\documentclass[journal ]{new-aiaa}
\usepackage[utf8]{inputenc}
\usepackage{textcomp}
\usepackage[tight]{subfigure} 
\usepackage{graphicx}
\usepackage{amsmath}
\usepackage[version=4]{mhchem}
\usepackage{siunitx}
\usepackage{longtable,tabularx}
\title{Influence of geometric structure, convection, and eddy on sound propagation in acoustic metamaterial with turbulent flow}

\author{Myong Chol Pak\footnote{Associate Professor, Department of Physics, Address/myongcholpak@163.com.},  Kwang-Il Kim\footnote{Associate Professor, Department of Physics, Address/kwangilkim@163.com.}, Hak Chol Pak\footnote{Professor, Department of Physics, Address/hcpak@163.com.}, and Kwon Ryong Hong\footnote{Research fellow, Institute of Natural Sciences, Address/hongkwonryong@163.com.}}
\affil{ Kim Il Sung University, Taesong District, 136, Pyongyang, Democratic People’s Republic of Korea}

\begin{document}

\maketitle

\begin{abstract}
The problem of reducing noise in the transportation is an important research field to prevent accidents and to provide a civilized environment for people.  A material that has recently attracted attention in research to reduce noise is acoustic metamaterial, and most of the research projects so far have been limited to the case of static media without flow. We have studied the sound transmission properties of acoustic metamaterial with turbulent flow to develop acoustic metamaterial that be used in transportation. In this paper, the effect of geometrical structure, the convective effect, and the eddy effect on sound propagation in acoustic metamaterial with turbulent flow are investigated, and the relationships between them are analyzed. The convective effect and the eddy effect both reduce the resonant strength of sound transmission loss resulting from the unique geometry of the acoustic crystal, but shift the resonant frequencies in opposite directions. In addition, when the convective effect and the eddy effect of the airflow, as well as the intrinsic interaction effect generated from the unique geometrical structure of the acoustic metamaterial cannot be ignored, they exhibit competition phenomena with each other, resulting in a widening of the resonance peak. As a result, these three effects cause the shift of the resonance frequency of the sound transmission loss and the widening of the resonance peak. The results of this study show that even in the case of turbulent flow, acoustic metamaterial can be used for transportation by properly controlling the geometric size and shape of the acoustic metamaterial.
\end{abstract}

\section*{Nomenclature}


{\renewcommand\arraystretch{1.0}
\noindent\begin{longtable*}{@{}l @{\quad=\quad} l@{}}
BLI & boundary-layer ingestion \\
CFD & computational field dynamics \\
$c_0$ & Speed of sound, m/s \\
$k$  & turbulent kinetic energy, m$^2/$s$^2$  \\
$k_0$  & wave number of incident acoustic wave, m$^{-1}$  \\
LBM & lattice Boltzmann method \\
Ma & Mach number \\
$L_p$ & sound pressure level, dB  \\
PML & Perfectly Matched Layer \\
$p_{rms}$ & root mean square pressure, Pa  \\
$p_{ref}$ & reference pressure for zero level corresponding to 0dB, Pa  \\
RANS &  Reynolds-averaged Navier-Stokes \\
$r_a$ & radius of annular cavity, m \\
$r_d$ & radius of circular duct, m \\
SPL &  sound pressure level \\
SST &  Shear Stress Transport \\
$t_a$ & height of annular cavity, m \\
$t_d$ & half of the height of neck in circular duct, m \\
TL & transmission loss, dB \\
$t_s$ & height of acoustic source, m \\
$t_p$ & height of PML, m \\
$\alpha_p$ &  coefficient of thermal expansion, K$^{-1}$ \\
$\beta_T$ &  isothermal compressibility, Pa$^{-1}$ \\
$\varepsilon$ &  turbulent dissipation rate, m$^2/$s$^3$ \\
$\eta$ &  Kolmogorov's scale, m \\
$\mu$ &  dynamic viscosity,  Pa$\cdot$s \\
$\mu_B$ &  bulk viscosity,  Pa$\cdot$s \\
$\mu_T$ &  turbulent dynamic viscosity,  Pa$\cdot$s \\
$\nu$ &  fluid kinematic viscosity,  m$^2$/s \\
$\nu_T$ &  turbulent kinematic viscosity,  m$^2$/s \\
$\tau$ &  viscous stress tensor, Pa \\
$\tau_T$ &  turbulence time scale, s \\
$\omega$ &  specific dissipation rate, s$^{-1}$ \\  
$\omega_0$ &  angular frequency of incident acoustic wave, s$^{-1}$ \\   
\end{longtable*}}

\section{Introduction}
\lettrine{W}{ith} the recent development of technology, attentions been focused on the improvement of human environment, and interests in noise reduction are increasing. Among them, acoustic metamaterial are widely applied because they can reduce noise by controlling the density and the bulk modulus of the material.

Studies have been discussed using acoustic metamaterial to control sound transmission by absorbing low-frequency sound in linear and nonlinear regions \cite{1brookea2020, 2Li2016} and doping impurities inside zero-index metamaterial \cite{3Gu2020}. Also, the  method for minimizing indoor sound energy by using an acoustic metamaterial with a flat panel structure \cite{4Qu2020} and  the method for detecting an acoustic image by constructing an acoustic superlens using membrane-based a two-dimensional metamaterial  having a negative density were reported \cite{5Park2015}. In addition, a non-resonant metasurface design for broadband elastic wave mode division that be used for elastic communication and biomedical diagnostic evaluation has also been proposed \cite{6Zheng2020}. Lu K. et al. showed through simulation that acoustic metamaterial with honeycomb structure effectively cause acoustic transmission loss in the low frequency range \cite{7Lu2016} and Fan L. et al. proved that the plates with circular holes blocked by membranes are effective in sound insulation at low frequencies by using numerical analysis \cite{8Fan2015}. Wang X.et al. proposed that a sound insulation effect can be obtained in the low frequency range by controlling the shape, stiffness and position of a thin film-type acoustic metamaterial with a stick fixed in the middle of the frame \cite{9Wang2016}. Acoustic metamaterial used to block broadband noise including low-frequency regions in air can be applied to water as well as air. Bok E.et al. proposed a way to use the acoustic metasurface consisting of a membrane and an air cavity filled with meta atoms in order to increase the acoustic sensitivity in water \cite{10Bok2018}.

As the practical applicability of acoustic metamaterial increases, research projects on acoustic metamaterial panels that can increase the noise reduction function while passing through a fluid are actively taking place \cite{11Su2014,12Jiang2017,13Sui2015}. In Ref. \cite{14Jung2018}, they designed an acoustic metamaterial panel that does not interfere with the flow of fluid while reducing the noise in a broadband in the audible frequency range. The proposed acoustic metamaterial panel allows the fluid to pass through the straight hole, but serves to block broadband noise by periodic annular cavities surrounding the hole. However, in these papers, the effect of the flow velocity of the fluid passing through the acoustic metamaterial on the sound wave is not discussed.

Meanwhile, research projects to control sound wave propagation in laminar and turbulent flows of fluids are also attracting attention. Yang Z. et al. proposed the idea that sound waves can propagate in one direction along the surface regardless of the presence of defects or obstacles in the acoustic structure with laminar circulation flow \cite{15Yang2015}. Research projects for investigating sound propagation properties in turbulent flow rather than laminar flow are attracting a lot of attention because they have a lot to do with practical applications. Turbulence effect of the fuselage on the fan noise of the BLI (Boundary-Layer Ingestion) configuration \cite{16Romani2020}, the relationship between the structural flexibility of the elastic trailing-edge and the aeroacoustic response \cite{17Nardini2020}, prediction of broadband sound generated in a low Mach number turbulent boundary layer by lattice Boltzmann method (LBM) \cite{18Kusano2020}, simulation of indoor noise generated by indoor window vibration \cite{19Yao2019}, and acoustic source model to reduce aerodynamic noise generated from wind turbine blades \cite{20Tang2019} have been discussed. Most of the interest, such as the reduction of aerofoil interaction noise by new serration profile group \cite{21Chaitanya2020,22Miotto2018}, the noise generation mechanism of controlled diffusion aerofoil and their dependence on Mach number \cite{23Deuse2020}, and the role of the porous material placed on the tail edge of the 3D chamber airfoil \cite{24Ananthan2020}, focuses on the reduction of noise caused by the interaction between aerofoil and turbulent flow.

Most of the researchers are only interested in sound wave control and noise generation in turbulent flows, but there are few studies on the effect of geometric structure, convective, and eddy effect on sound propagation in acoustic metamaterial with turbulent flow. Therefore, we discuss the convective and eddy effects on acoustic propagation as turbulence flows into the acoustic metamaterial consisting of straight holes and periodic annular cavities surrounding the hole. Also, in this case, the change in broadband acoustic wave blocking characteristics according to the geometric size and the number of annular cavities is investigated. This paper is organized as follows. Section 2 describes the theoretical basis for aeroacoustic properties in turbulent flows. In Section 3, numerical results for sound transmission loss and sound pressure level of acoustic metamaterial are shown and analyzed in both cases of no flow and turbulent flow. In particular, the turbulence flowing in the acoustic metamaterial is analyzed by CFD (computational fluid dynamics), and based on the results, the convective effect and the eddy effect on sound transmission are discussed. Also, the sound transmission properties according to the geometric size of acoustic crystals and the number of ring-shaped cavities are also considered. Finally, in Section 4, the influence of geometric structure, convection, and eddy on sound propagation in acoustic metamaterial with turbulent flows are concluded, and future application prospects are described.

\section{Theoretical Background}

Using the linearized Navier-Stokes equation, we study the propagation properties of sound waves in a fluid. This equation consists of the continuity, momentum, and energy equations \cite{25Ostashev2016book}.

\begin{equation}
\label{eq1}
\frac{{\partial {\rho _t}}}{{\partial t}} + \nabla  \cdot ({\rho _0}{{\bf{u}}_t} + {\rho _t}{{\bf{u}}_0}) = M
\end{equation}
\begin{equation}
\label{eq2}
{\rho _0}[\frac{{\partial {{\bf{u}}_t}}}{{\partial t}} + ({{\bf{u}}_t} \cdot \nabla ){{\bf{u}}_0} + ({{\bf{u}}_0} \cdot \nabla ){{\bf{u}}_t}] + {\rho _t}({{\bf{u}}_0} \cdot \nabla ){{\bf{u}}_0} = \nabla  \cdot {\bf{\sigma }} + {\bf{F}} - {{\bf{u}}_0}M
\end{equation}
\begin{equation}
\label{eq3}
\begin{array}{l}
{\rho _0}{C_p}[\frac{{\partial {T_t}}}{{\partial t}} + ({{\bf{u}}_t} \cdot \nabla ){T_0} + ({{\bf{u}}_0} \cdot \nabla ){T_t}] + {\rho _t}{C_p}({{\bf{u}}_0} \cdot \nabla ){T_0}\\
- {\alpha _p}{T_0}[\frac{{\partial {p_t}}}{{\partial t}} + ({{\bf{u}}_t} \cdot \nabla ){p_0} + ({{\bf{u}}_0} \cdot \nabla ){p_t}] - {\alpha _p}{T_t}({{\bf{u}}_0} \cdot \nabla ){p_0} = \nabla  \cdot (\kappa \nabla {T_t}) + \Phi  + Q
\end{array}
\end{equation}
where ${p_t}$, ${{\bf{u}}_t}$, ${T_t}$, and ${\rho_t}$ are the acoustic perturbations to the pressure, the velocity, the temperature, and the density, respectively. ${p_t}$, ${{\bf{u}}_t}$, and ${T_t}$ are equal to the sum of the physical quantities in the background acoustic field and the scattered field.

\begin{equation}
\label{eq4}
{p_t} = p + {p_b},\;{{\bf{u}}_t} = {\bf{u}} + {{\bf{u}}_b},\;{T_t} = T + {T_b}
\end{equation}
Also, $M$, ${\bf{F}}$, $Q$, $C_p$, $\alpha_p$, and $\kappa$ are the mass source, the volume force source, the volumetric heat source, the heat capacity at constant pressure, the coefficient of thermal expansion, and the thermal conductivity, respectively. Additionally, the stress tensor, the linearized equation of state and the linearized viscous dissipation function are defined as,
\begin{equation}
\label{eq5}
{\bf{\sigma }} =  - {p_t}{\bf{I}} + \mu (\nabla {{\bf{u}}_t} + {(\nabla {{\bf{u}}_t})^T}) + ({\mu _B} - 2\mu /3)(\nabla  \cdot {{\bf{u}}_t}){\bf{{\rm I}}}
\end{equation}
\begin{equation}
\label{eq6}
{\rho _t} = {\rho _0}({\beta _T}{p_t} - {\alpha _p}{T_t})
\end{equation}
\begin{equation}
\label{eq7}
\Phi  = \nabla {{\bf{u}}_t}:{\bf{\tau }}({{\bf{u}}_0}) + \nabla {{\bf{u}}_0}:{\bf{\tau }}({{\bf{u}}_t})
\end{equation}
\begin{equation}
\label{eq8}
{\bf{\tau }}({{\bf{u}}_t}) = \mu [\nabla {{\bf{u}}_t} + {(\nabla {{\bf{u}}_t})^T}] + ({\mu _B} - 2\mu /3)(\nabla  \cdot {{\bf{u}}_t}){\bf{{\rm I}}}
\end{equation}
\begin{equation}
\label{eq9}
{\bf{\tau }}({{\bf{u}}_0}) = \mu [\nabla {{\bf{u}}_0} + {(\nabla {{\bf{u}}_0})^T}] + ({\mu _B} - 2\mu /3)(\nabla  \cdot {{\bf{u}}_0}){\bf{{\rm I}}}
\end{equation}
where $\bf{\tau }$, $\beta_T$, $\mu$, and $\mu_B$ are the viscous stress tensor, the isothermal compressibility, the dynamic viscosity and the bulk viscosity,  respectively.  In the linearized Navier-Stokes equation, $p_0$, ${\bf{u}}_0$, $T_0$, and $\rho_0$ are absolute pressure, velocity, temperature, and density of the background mean flow used to account for the effect of the background mean flow on the sound wave. This is calculated by using the CFD study of the fluid.

When sound waves propagate into a turbulent flow, the flow properties are evaluated by Reynolds-averaged Navier-Stokes (RANS) model \cite{26Menter1994}.
The Reynolds-averaged representation of turbulent flows divides the flow quantities into a time-averaged part and a fluctuating part.
\begin{equation}
\label{eq10}
{{\bf{u}}_0} = {{\bf{\bar u}}_0} + {{\bf{u'}}_0},\,{\rho _0} = {\bar \rho _0} + {\rho '_0},\;{p_0} = {\bar p_0} + {p'_0}
\end{equation}
In order to discuss the turbulent flow, the SST (Shear Stress Transport) turbulent method is used among various RANS models \cite{25Ostashev2016book}. The advantage of this method is that it can describe the flow characteristics well close to the wall and the dependence on the initial parameters of the main free stream flow is not very large.

The basic equation governed is as follows.
\begin{equation}
\label{eq11}
\frac{{\partial {{\bar \rho }_0}}}{{\partial t}} + \nabla  \cdot ({\bar \rho _0}{{\bf{\bar u}}_0}) = 0
\end{equation}
\begin{equation}
\label{eq12}
\begin{array}{l}
{{\bar \rho }_0}\frac{{\partial {{{\bf{\bar u}}}_0}}}{{\partial t}} + {{\bar \rho }_0}({{{\bf{\bar u}}}_0} \cdot \nabla ){{{\bf{\bar u}}}_0} = \nabla  \cdot \{  - {{\bar p}_0}{\bf{I}} + (\mu  + {\mu _T})[\nabla {{{\bf{\bar u}}}_0} + {(\nabla {{{\bf{\bar u}}}_0})^T}]\\
\;\;\;\;\;\;\;\;\;\;\;\;\; - \frac{2}{3}(\mu  + {\mu _T})(\nabla  \cdot {{{\bf{\bar u}}}_0}){\bf{I}} - \frac{2}{3}{{\bar \rho }_0}k{\bf{I}}\}  + {\bf{F}}
\end{array}
\end{equation}
The model equations are formulated in the averaged turbulent kinetic energy $k$ and the turbulent frequency $\omega$, 
\begin{equation}
\label{eq13}
{\bar \rho _0}\frac{{\partial k}}{{\partial t}} + {\bar \rho _0}({{\bf{\bar u}}_0} \cdot \nabla )k = \nabla  \cdot [(\mu  + {\mu _T}{\sigma _k})\nabla k] + P - {\bar \rho _0}\beta _0^*k\omega
\end{equation}
\begin{equation}
\label{eq14}
{\bar \rho _0}\frac{{\partial \omega }}{{\partial t}} + {\bar \rho _0}({{\bf{\bar u}}_0} \cdot \nabla )\omega  = \frac{{{{\bar \rho }_0}\gamma }}{{{\mu _T}}}P - {\bar \rho _0}\beta {\omega ^2} + \nabla  \cdot [(\mu  + {\mu _T}{\sigma _\omega })\nabla \omega ] + 2(1 - {f_{v1}})\frac{{{{\bar \rho }_0}{\sigma _{\omega 2}}}}{\omega }\nabla \omega  \cdot \nabla k
\end{equation}
where,
\begin{equation}
\label{eq15}
P = \min ({P_k},\;10{\bar \rho _0}\beta _0^*k\omega )
\end{equation}
\begin{equation}
\label{eq16}
{P_k} = {\mu _T}\{ \nabla {{\bf{\bar u}}_0}:[\nabla {{\bf{\bar u}}_0} + {(\nabla {{\bf{\bar u}}_0})^T}] - \frac{2}{3}{(\nabla  \cdot {{\bf{\bar u}}_0})^2}\}  - \frac{2}{3}{\bar \rho _0}k\nabla  \cdot {{\bf{\bar u}}_0}
\end{equation}
In this case, the turbulent eddy viscosity is,
\begin{equation}
\label{eq17}
{\mu _T} = \frac{{{{\bar \rho }_0}{\alpha _1}k}}{{\max ({\alpha _1}\omega ,S{f_{v2}})}}
\end{equation}
where $S$ is the characteristic magnitude of the mean velocity gradients,
\begin{equation}
\label{eq18}
S = \sqrt {2S_{ij}S_{ij}}
\end{equation}
and $S_{ij}$ is the mean strain-rate tensor,
\begin{equation}
\label{eq19}
{S_{ij}} = \frac{1}{2}(\frac{{\partial {{\bar u}_{0i}}}}{{\partial {x_j}}} + \frac{{\partial {{\bar u}_{0j}}}}{{\partial {x_i}}})
\end{equation}
The constants $\beta$, $\gamma$, $\sigma_k$, and $\sigma_\omega$ are interpolated values between inner and outer values.
\begin{equation}
\label{eq20}
\left\{ \begin{array}{l}
\beta  = {f_{v1}}{\beta _1} + (1 - {f_{v1}}){\beta _2}\\
\gamma  = {f_{v1}}{\gamma _1} + (1 - {f_{v1}}){\gamma _2}\\
{\sigma _k} = {f_{v1}}{\sigma _{k1}} + (1 - {f_{v1}}){\sigma _{k2}}\\
{\sigma _\omega } = {f_{v1}}{\sigma _{\omega 1}} + (1 - {f_{v1}}){\sigma _{\omega 2}}
\end{array} \right.
\end{equation}
The interpolation functions $f_{v1}$ and $f_{v2}$  are
\begin{equation}
\label{eq21}
{f_{v1}} = \tanh (\theta _1^4)
\end{equation}
and,
\begin{equation}
\label{eq22}
{f_{v2}} = \tanh (\theta_2^2).
\end{equation}
In this case,
\begin{equation}
\label{eq23}
{\theta _1} = \min [\max (\frac{{\sqrt k }}{{\beta _0^*\omega {l_\omega }}},\;\frac{{500\mu }}{{{{\bar \rho }_0}\omega l_\omega ^2}}),\,\frac{{4{{\bar \rho }_0}{\sigma _{\omega 2}}k}}{{C{D_{k\omega }}l_\omega ^2}}],
\end{equation}
\begin{equation}
\label{eq24}
C{D_{k\omega }} = \max (\frac{{2{{\bar \rho }_0}{\sigma _{\omega 2}}}}{\omega }\nabla \omega  \cdot \nabla k,\;{10^{ - 10}}),
\end{equation}
\begin{equation}
\label{eq25}
{\theta _2} = \max (\frac{{2\sqrt k }}{{\beta _0^*\omega {l_\omega }}},\;\frac{{500\mu }}{{{{\bar \rho }_0}\omega l_\omega ^2}}).
\end{equation}
where $\beta_1=0.075$, $\beta_2=0.0828$, $\gamma_1=5/9$, $\gamma_2=0.44$, $\sigma_{k1}=0.85$, $\sigma_{k2}=1$, $\sigma_{\omega_1}=0.5$, $\sigma_{\omega_2}=0.856$, $\alpha_1=0.31$,  $\beta _0^*=0.09$, and $l_\omega$ is the distance to the closest wall.

\section{Results and Analysis}

In this section, we first describe the simulation parameters and acoustic characteristic parameters of the acoustic metamaterial to be considered. Also, in order to investigate the effect of the geometrical structure on sound transmission, sound pressure level characteristic values and transmission loss results are calculated and analyzed using finite element simulation in the case of no flow. Next, when there is turbulent flow, the  CFD analysis of the flow is performed, and sound transmission properties are investigated for the flow velocity, and the convective effect and the eddy effect are compared with each other. Finally, while changing the geometric structure parameter, we observe the change of sound pressure level and sound transmission loss values.

\subsection{Simulation parameters}
Fig. 1 is a design diagram of an acoustic metamaterial consisting of a straight hole through which a fluid can pass and periodic annular cavities surrounding the hole.
\begin{figure}[hbt!]
	\centering
	\includegraphics[width=.5\textwidth]{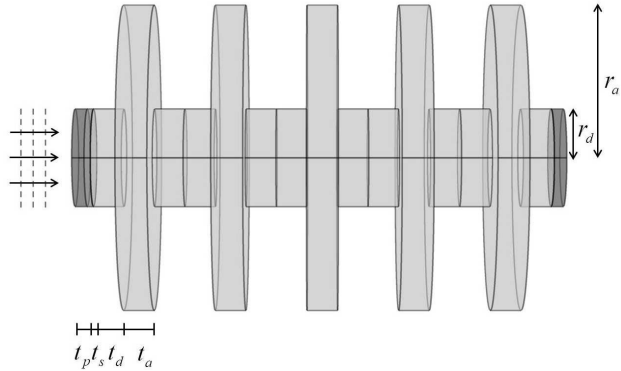}
	\caption{A design diagram of the acoustic metamaterial to be discussed.}
\end{figure}
In Fig.1, $r_a$ is the radius of annular cavity, $r_d$ is the radius of circular duct, $t_a$ is the height of annular cavity, $t_d$ is a half of the height of neck in circular duct, $t_s$ is the height of acoustic source, and $t_p$ is the height of PML(Perfectly Matched Layer). Table 1 shows the remaining geometric structure parameter values excluding $r_a$  because the calculation is performed while changing the size of the radius $r_a$.

\begin{table}[hbt!]
	\caption{\label{tab:table1} Geometric structure parameters}
	\centering
	\begin{tabular}{ccccc}
		\hline
		$r_d$(mm)& $t_a$(mm)& $t_d$(mm)& $t_s$(mm)& $t_p$(mm)\\\hline
		8& 5& 5& 1& 2\\
		\hline
	\end{tabular}
\end{table}

As shown in Fig. 1, the discussed acoustic wave is a plane wave, and its incident surface is perpendicular to the rotation axis direction of the acoustic metamaterial. The finite element simulation was performed with the commercial software COMSOL Multiphysics 5.5. In this case, the boundary condition of the numerical simulation is assumed to be no slip and the sound velocity is ${c_0} = 343\;{\rm{m/s}}$. PML is applied at the inlet and outlet, which acts to absorb acoustic waves by simulating an open boundary. We calculated the sound pressure in the frequency range of 2000 to 6000 Hz and evaluated the transmission loss and sound pressure level based on it. In this case, the transmission loss of the system is defined as,
\begin{equation}
\label{eq26}
TL = 20{\log _{10}}(\left| {\frac{{{p_{in}}}}{{{p_{out}}}}} \right|)
\end{equation}
where $p_{in}$ and $p_{out}$  are the average pressure at the inlet and outlet, respectively \cite{27Du2016,28Gikadi2014}.  And when the sound pressure $p$ changes harmonically with time, the sound pressure level (SPL) $L_p$ is expressed by the root mean square (rms) pressure $p_{rms}$ of the fluid, such as
\begin{equation}
\label{eq27}
{L_p} = 20{\log _{10}}(\frac{{{p_{rms}}}}{{{p_{ref}}}})\;,\;\;{p_{rms}} = \sqrt {p{p^*}/2}
\end{equation} 
where $p_{ref}$ is the reference pressure for the zero level corresponding to 0dB \cite{29Pierce2019book}. The zero level corresponding to this dB scale depends on the type of fluid. For example, the reference pressure for air is 20$\mu$Pa and the reference pressure for water is 1$\mu$Pa .

\subsection{Calculation results of sound pressure level in case of no flow}

We first investigate the acoustic pressure of the acoustic metamaterial in case of no flow in order to evaluate the effect of the geometry on the sound transmission of the acoustic metamaterial with turbulent flow. In this case, an incident acoustic plane wave with an amplitude of 1Pa is incident in a direction perpendicular to the axis of rotation of the acoustic metamaterial in the area marked in red in Fig. 1. Since the properties of the sound pressure vary according to the geometric size, we investigated the change in the sound pressure level corresponding to the radius of annular cavity and the number of annular cavities.

Fig. 2(a) shows the sound pressure level values versus the frequencies of the sound wave when there is no flow (Ma=0) and the radius of annular cavity is 22.5mm. As shown in Fig. 2(a), as the frequency increases initially, the sound pressure level values decrease rapidly, have a minimum value at a specific frequency (4529.1 Hz in the case of N=5), and then rise again. In other words, it has a resonant property, which can be treated as the result of the interaction between the particle vibration in the direction of sound propagation in the circular duct and the particle vibration perpendicular to the direction of sound propagation in the annular cavity \cite{14Jung2018}. 

\begin{figure}[!htb]
	\centering
	\subfigure[$r_a$=22.5mm, Ma=0]
	{ \includegraphics[scale=0.65]{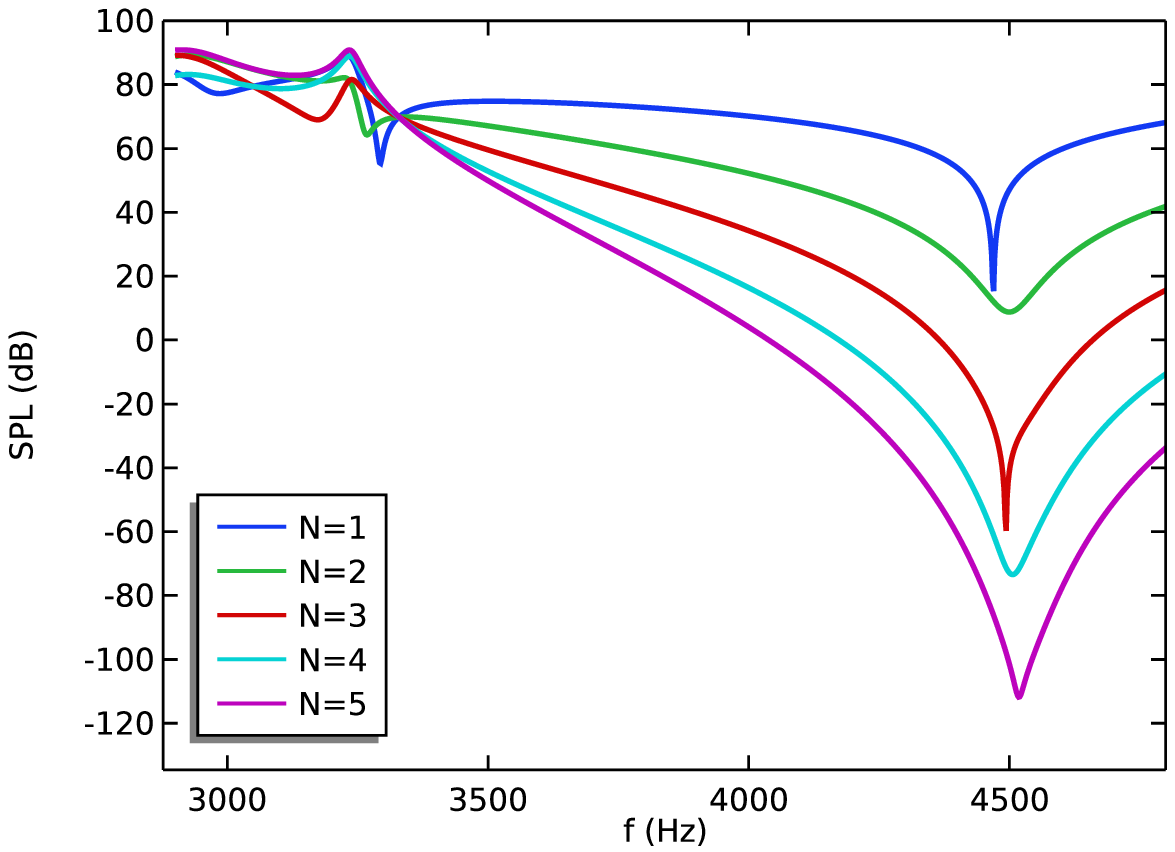}
		\label{SPL225a}
	}\
	\subfigure[$r_a$=25.0mm, Ma=0]
	{ \includegraphics[scale=0.65]{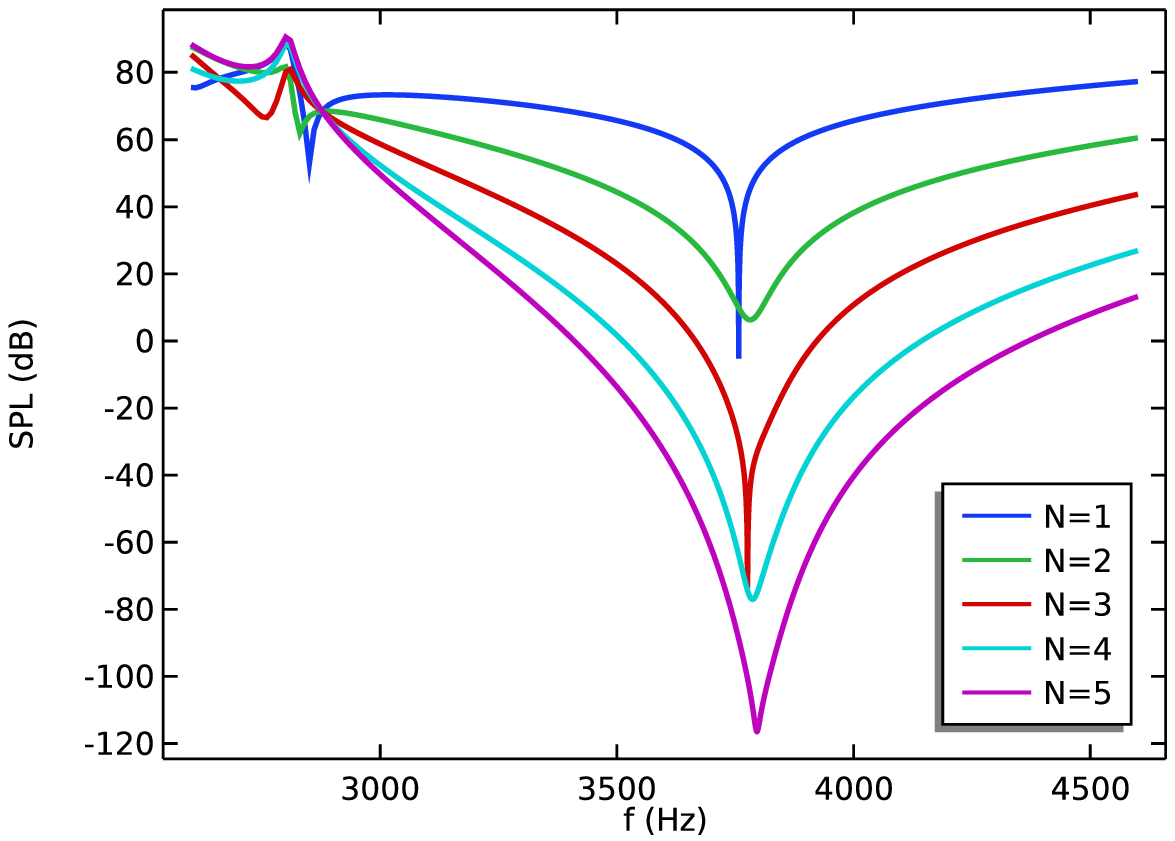}
		\label{SPL250b}
	}\\
	\subfigure[$r_a$=27.5mm, Ma=0]
	{ \includegraphics[scale=0.65]{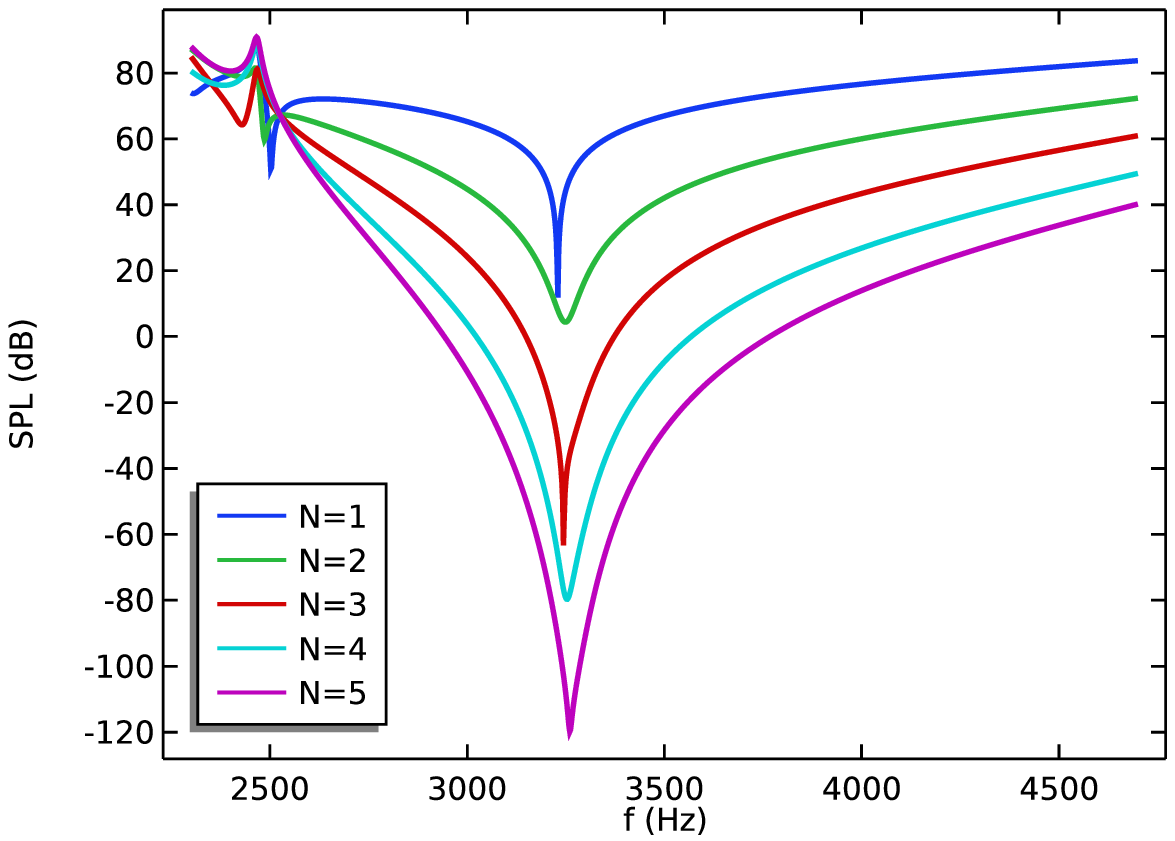}
		\label{SPL275c}
	}
	\subfigure[$r_a$=30.0mm, Ma=0]
	{ \includegraphics[scale=0.65]{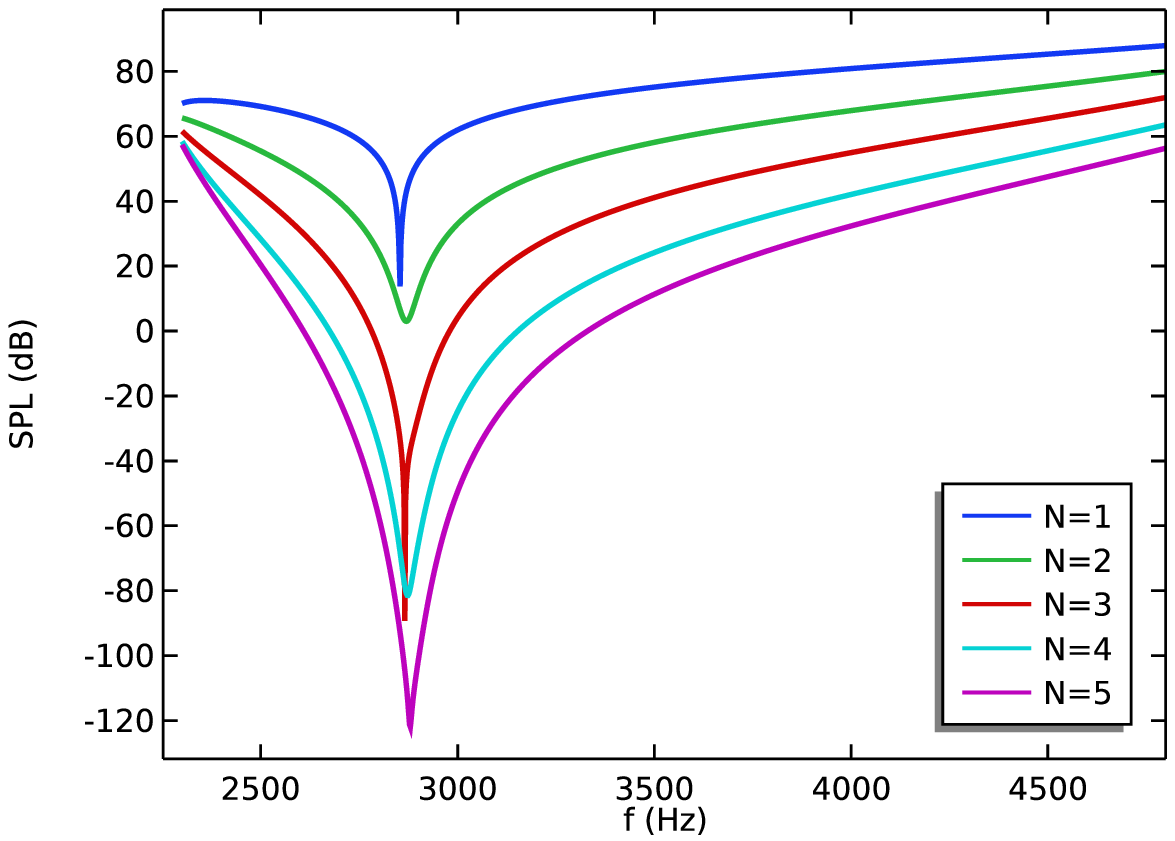}
		\label{SPL300d}
	}
	\caption{ In the case of no flow, the change in sound pressure level according to the change in the radius of annular cavity and number of annular cavities.}
\end{figure}
This property appears regardless of the number of annular cavities. However, as the number increases, the magnitude of these minima becomes obviously smaller and the resonance strength becomes stronger. This is because, as the number of annular cavities increases, the cross-region between the particle vibration in the sound propagation direction in the circular duct and the particle vibration perpendicular to the sound propagation direction in the annular cavity increases, and as a result, the interaction becomes stronger. In the case of N=1, another peak appears at 3293.8Hz. Its peak intensity is smaller than the peak discussed earlier, and it gets weaker as the number of annular cavities increases. This is a peak related to the length of the circular duct. In the case of N=1, this peak cannot be ignored because the resonance related to the interaction is not very large. However, as the number of annular cavities increases, the interaction between the particle vibration in the direction of sound propagation in the circular duct and the particle vibration perpendicular to the direction of sound propagation in the annular cavity increases, and the intensity of this peak disappears.

Even when $r_a$ are 25.0mm, 27.5mm, and 30.0mm, the resonance properties analyzed above still appear (Fig. 2(b), 2(c), 2(d)). Table 2 shows the resonant frequency values corresponding to the annular cavity radii when N=5. Note in Fig. 2 and Table 2 that the resonant frequency decreases as the radius of the annular cavity increases. This is analyzed with particle vibrations perpendicular to the direction of sound propagation in the annular cavity. As the radius of the annular cavity increases, the wavelength of the stationary wave in the cavity increases, and thus the resonance frequency decreases. As a result, the resonance frequency of the sound propagation is also lowered due to the interaction between the particle vibration in the sound propagation direction in the circular duct and the particle vibration perpendicular to the sound propagation direction in the annular cavity. This proves once again that this resonance property is closely related to the particle vibration in the annular cavity.

\begin{table}[hbt!]
	\caption{\label{tab:table2} Resonant frequency values corresponding to the radii of annular cavity in case of N=5}
	\centering
	\begin{tabular}{ccccc}
		\hline
		$r_a$(mm)& 22.5& 25.0& 27.5& 30.0\\\hline
		$f^{res}$(Hz)& 4529.1& 3804.8& 3267.6& 2862.3\\
		\hline
	\end{tabular}
\end{table}

\subsection{CFD analysis for turbulent flow}

In order to investigate the sound propagation properties of acoustic metamaterial with turbulent flow, CFD analysis of turbulence in airflow is performed using the SST model mentioned above. The velocity of turbulent flow is evaluated with the Mach number Ma, and the properties of the case where Ma is 0.02, 0.05, 0.10, 0.15, 0.17, 0.20, 0.22 are discussed. In this case, the kinematic viscosity of air is $\nu  = {\rm{1}}{\rm{.50}} \times {\rm{1}}{{\rm{0}}^{{\rm{ - 5}}}}\;{{\rm{m}}^{\rm{2}}}{\rm{/s}}$.

To investigate turbulent flow, the turbulent kinetic energy $k$, specific dissipation rate $\omega$, turbulent dissipation rate $\varepsilon$, turbulent dynamic viscosity $\mu_T$, turbulent kinematic viscosity $\nu_T$, and turbulence time scale $\tau_T$ should be evaluated and analyzed. Table 3 shows the turbulent flow parameters obtained at the outlet of the acoustic metamaterial by simulating while changing the velocity of the turbulent flow. As Ma increases, turbulent kinetic energy, specific dissipation rate, turbulent dissipation rate, turbulent dynamic viscosity, and turbulent kinematic viscosity increase. However, the turbulence time scale decreases as Ma increases, which is in good agreement with the fact that $\tau_T$ is inversely proportional to the specific dissipation rate $\omega$. Fig. 3 shows a quarter cross section of turbulent dynamic viscosity $\mu_T$ in the case of Ma = 0.15. It is evident that the turbulent dynamic viscosity is not zero in the annular cavity region of Fig.3, and this fact intuitively shows that the annular cavity region has a great influence on the airflow flowing through the circular duct.

\begin{table}[hbt!]
	\caption{\label{tab:table3} Turbulent flow parameters at the outlet of acoustic metamaterial}
	\centering
	\begin{tabular}{cccccccc}
		\hline
		Ma& 0.02& 0.05& 0.10& 0.15& 0.17& 0.20& 0.22\\\hline
		$k\;({{\rm{m}}^{\rm{2}}}{\rm{/}}{{\rm{s}}^{\rm{2}}})$& 0.00538&	0.225&	3.86&	14.6&	21.1&	33.3&	43.0\\\hline
		$\omega \;({10^5}{{\rm{s}}^{ - 1}})$& 8.92&	8.94&	9.25&	10.1&	10.5&	11.3&	11.9\\\hline
		$\varepsilon ({10^4}{{\rm{m}}^{\rm{2}}}{\rm{/}}{{\rm{s}}^{\rm{3}}})$& 0.0432&	1.81&	32.1&	132&	199&	338&	461\\\hline
		${\mu _T}({10^{ - 7}}{\rm{Pa}} \cdot {\rm{s}})$& 0.0727&	3.03&	50.2&	175&	242&	355&	436\\\hline
		${\nu _T}({10^{ - 7}}{{\rm{m}}^{\rm{2}}}{\rm{/s}})$& 0.0603&	2.52&	41.7&	145&	201&	294&	361\\\hline
		${\tau _T}({10^{ - 5}}{\rm{s}})$& 1.25&	1.24&	1.20&	1.11&	1.06&	0.983&	0.933\\\hline
	\end{tabular}
\end{table}

\begin{figure}[hbt!]
	\centering
	\includegraphics[width=.65\textwidth]{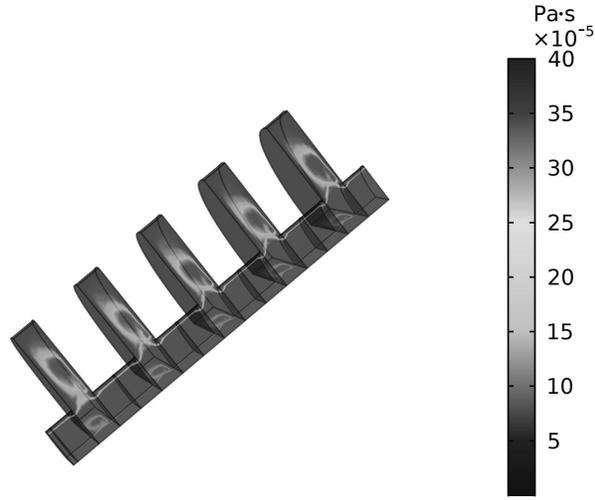}
	\caption{ Quarter cross section of the turbulent dynamic viscosity in the case of Ma=0.15.}
\end{figure}

It is important to evaluate the property of the turbulent dissipation rate $\varepsilon$ well in investigating the turbulent flow of a fluid. The turbulent dissipation rate $\varepsilon$ does not depend on the kinematic viscosity $\nu$, but instead is determined by the nature of the largest eddy that extracts energy from the mean flow \cite{30Kundu2012book}. A scale that well reflects the balance between the inertia effect and the viscous effect of eddy is Kolmogorov's scale, defined as,
\begin{equation}
\label{eq28}
\eta  = {({\nu ^3}/\varepsilon )^{1/4}}
\end{equation} 
Therefore, it is necessary to consider this scale carefully to evaluate the properties of turbulent flow and to analyze the sound transmission result accurately in turbulent flow. Thus, we investigated the Kolmogorov's scale versus the velocity of air (Fig. 4). As shown in Fig. 4, as Ma increases, Kolmogorov's scale gradually decreases. That is, the faster the velocity, the smaller the effect of eddy. In Fig. 4, the scale change was also investigated while changing the number of annular cavities, but there is no significant change depending on the number.

\begin{figure}[hbt!]
	\centering
	\includegraphics[width=.65\textwidth]{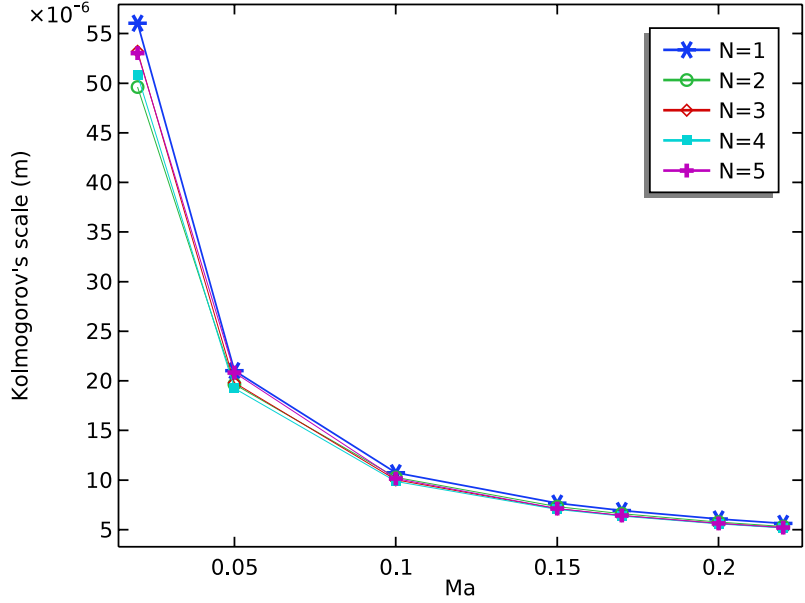}
	\caption{ Kolmogorov's scale versus the velocity of airflow.}
\end{figure}

\subsection{Sound transmission in turbulent flow}

We investigated the transmission properties of sound waves in turbulent flow with the linearized Navier-Stokes Model discussed in Section 2, based on determining the pressure, velocity, temperature, and density of turbulent flow in acoustic metamaterial. In this case, the incident acoustic plane wave of ${p_b} = {p_0}{e^{ - i{k_0}z}}$ is incident on the area marked in red in Fig. 1 in a direction perpendicular to the rotation axis of the acoustic metamaterial, where ${p_0} = 1{\rm{Pa}}$, ${k_0} = {\omega _0}/[{c_0}{\rm{(Ma}} + 1)]$, and $\omega_0$  is the angular frequencies of the incident acoustic wave.

\begin{figure}[hbt!]
	\centering
	\includegraphics[width=.65\textwidth]{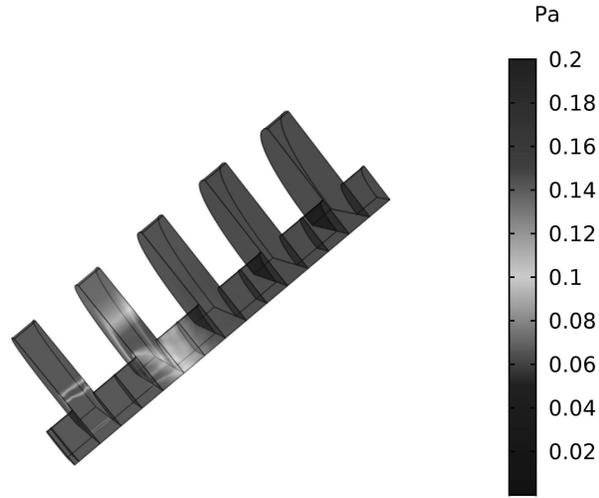}
	\caption{ A quarter cross section of the acoustic pressure in the case of $f$=5700Hz, Ma=0.15, and $r_a$=25mm.}
\end{figure}

Fig. 5 shows a quarter cross section of the acoustic pressure obtained as a simulation result for $f$=5700Hz, Ma=0.15, and $r_a$=25mm. This figure shows that even in turbulent flow, sound transmission can be relatively blocked due to the geometry of the acoustic metamaterial. Thus, we investigated the sound transmission loss while increasing the flow velocity.

\begin{figure}[!htb]
	\centering
	\subfigure[Ma=0.02, N=5]
	{ \includegraphics[scale=0.65]{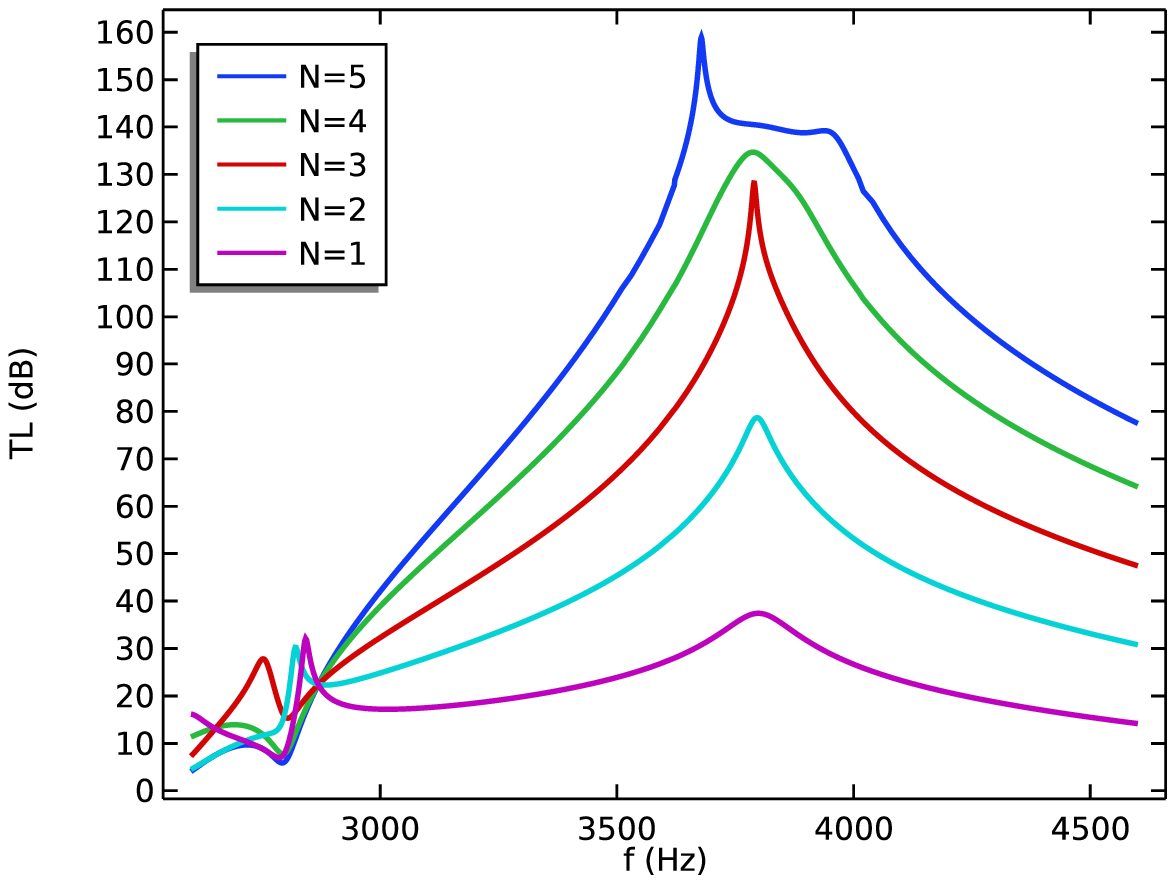}
		\label{TL002}
	}\
	\subfigure[Ma=0.05, N=5]
	{ \includegraphics[scale=0.65]{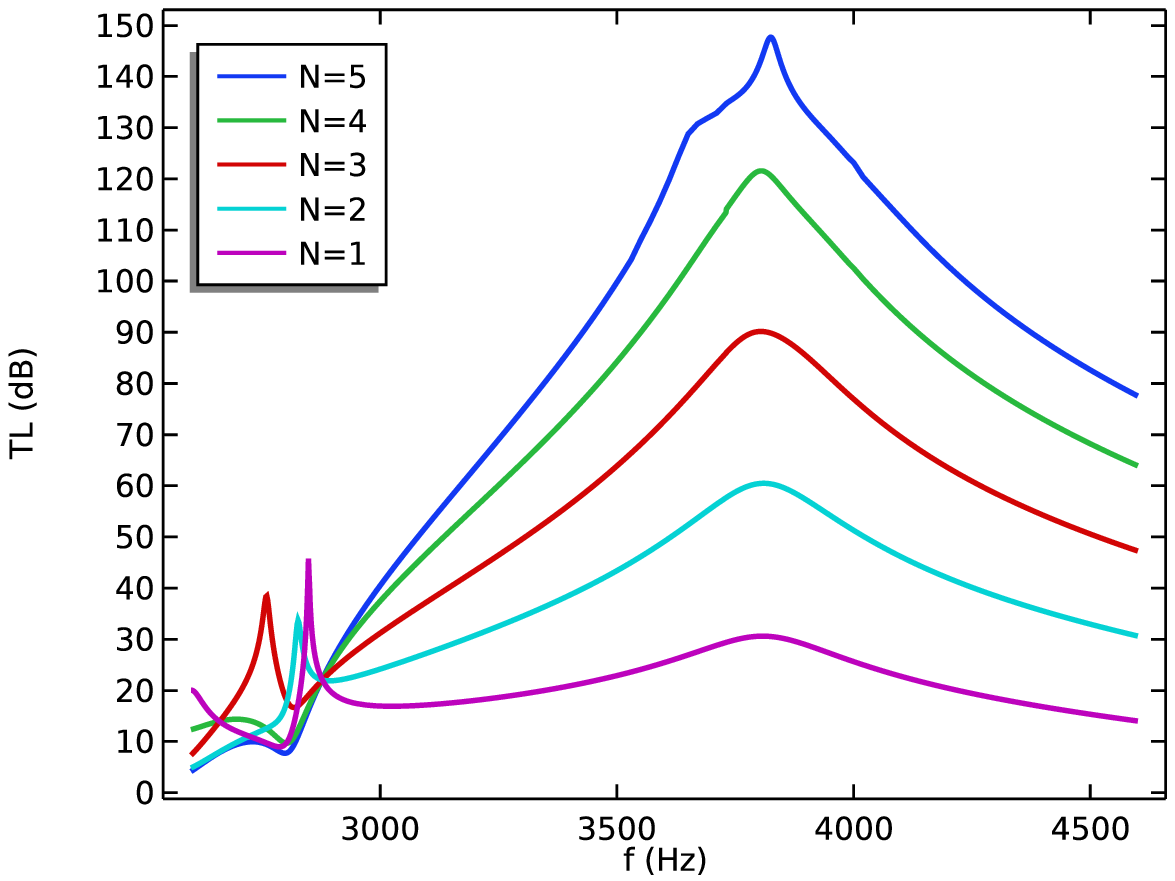}
		\label{TL005}
	}\
	\subfigure[Ma=0.10, N=5]
	{ \includegraphics[scale=0.65]{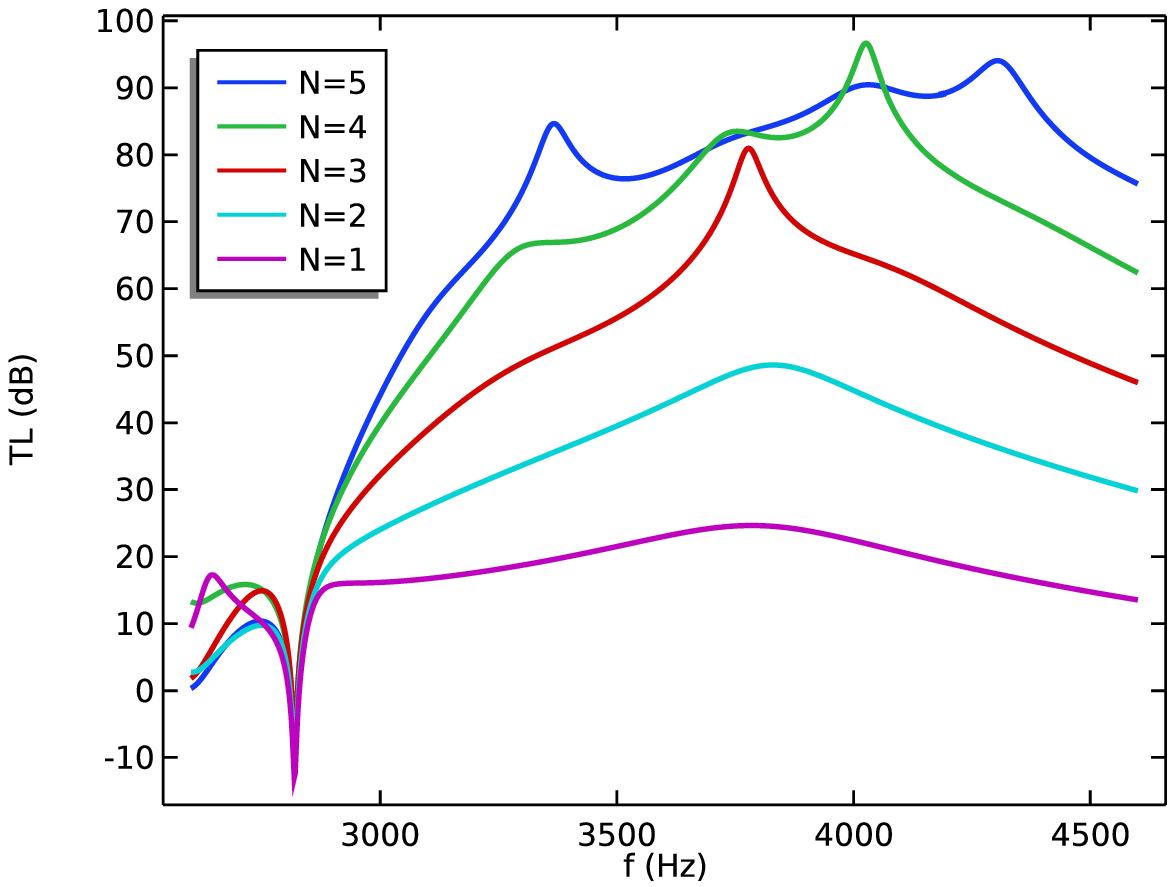}
		\label{TL010}
	}\
	\subfigure[Ma=0.15, N=5]
	{ \includegraphics[scale=0.65]{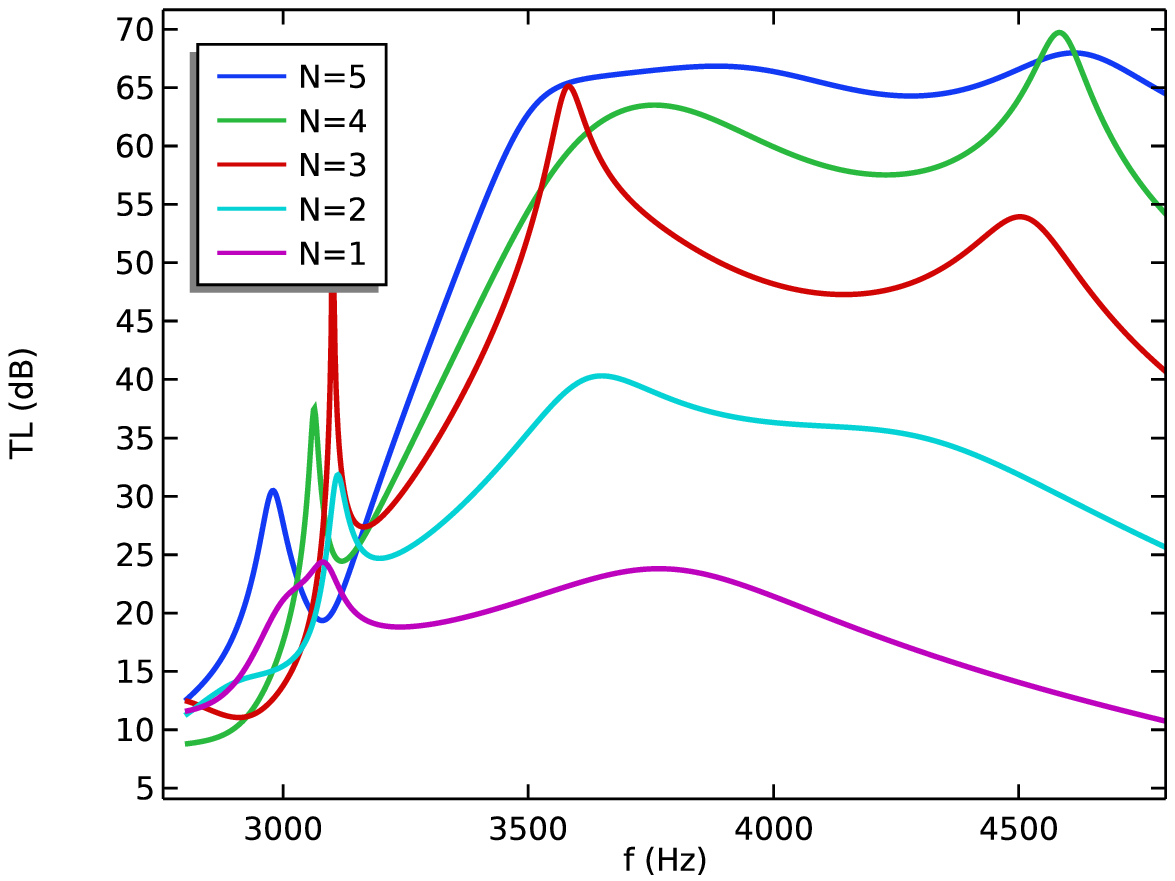}
		\label{TL015}
	}\

	\caption{ Sound transmission loss corresponding to the size of Ma and the number of annular cavities with $r_a$=25mm in the case of Ma=0.02, 0.05, 0.10, and 0.15.}
\end{figure}

\begin{figure}[!htb]
	\centering
	\subfigure[Ma=0.17, N=5]
	{ \includegraphics[scale=0.65]{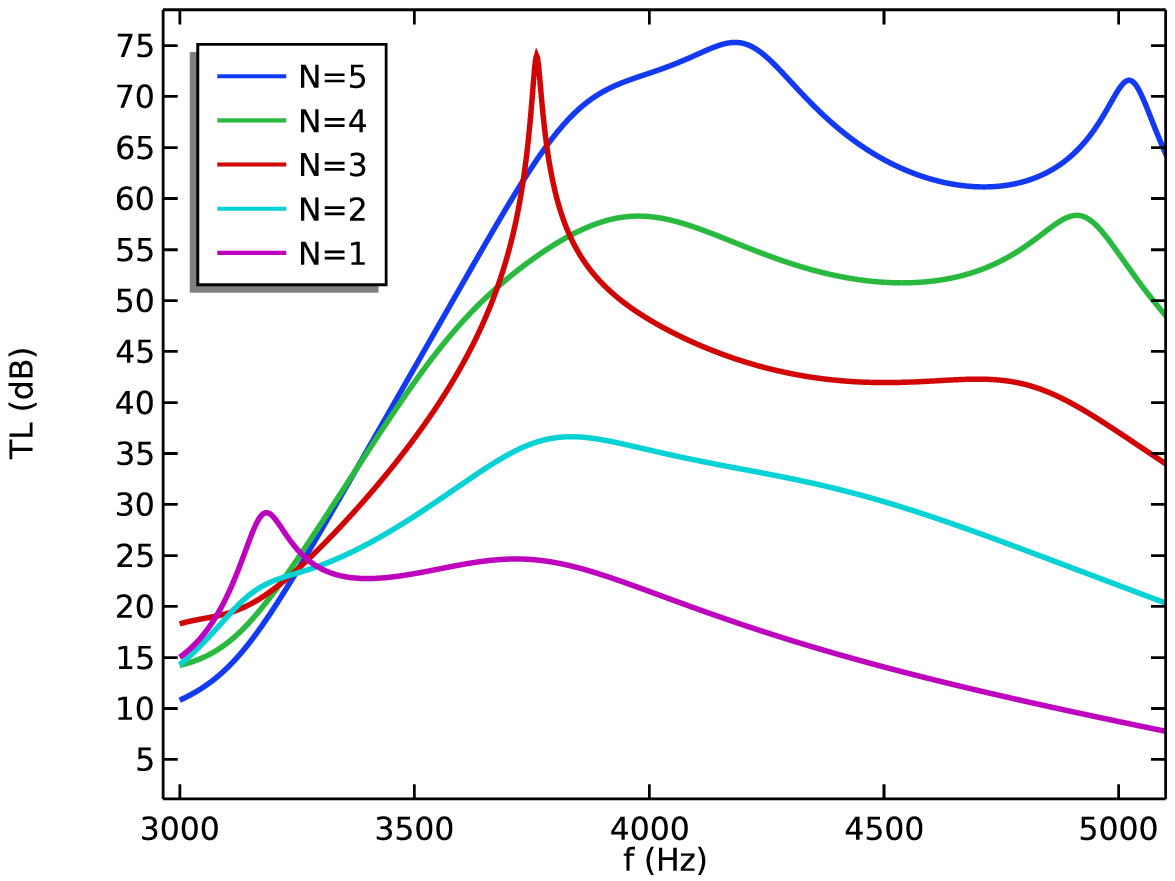}
		\label{TL017}
	}\
	\subfigure[Ma=0.20, N=5]
	{ \includegraphics[scale=0.65]{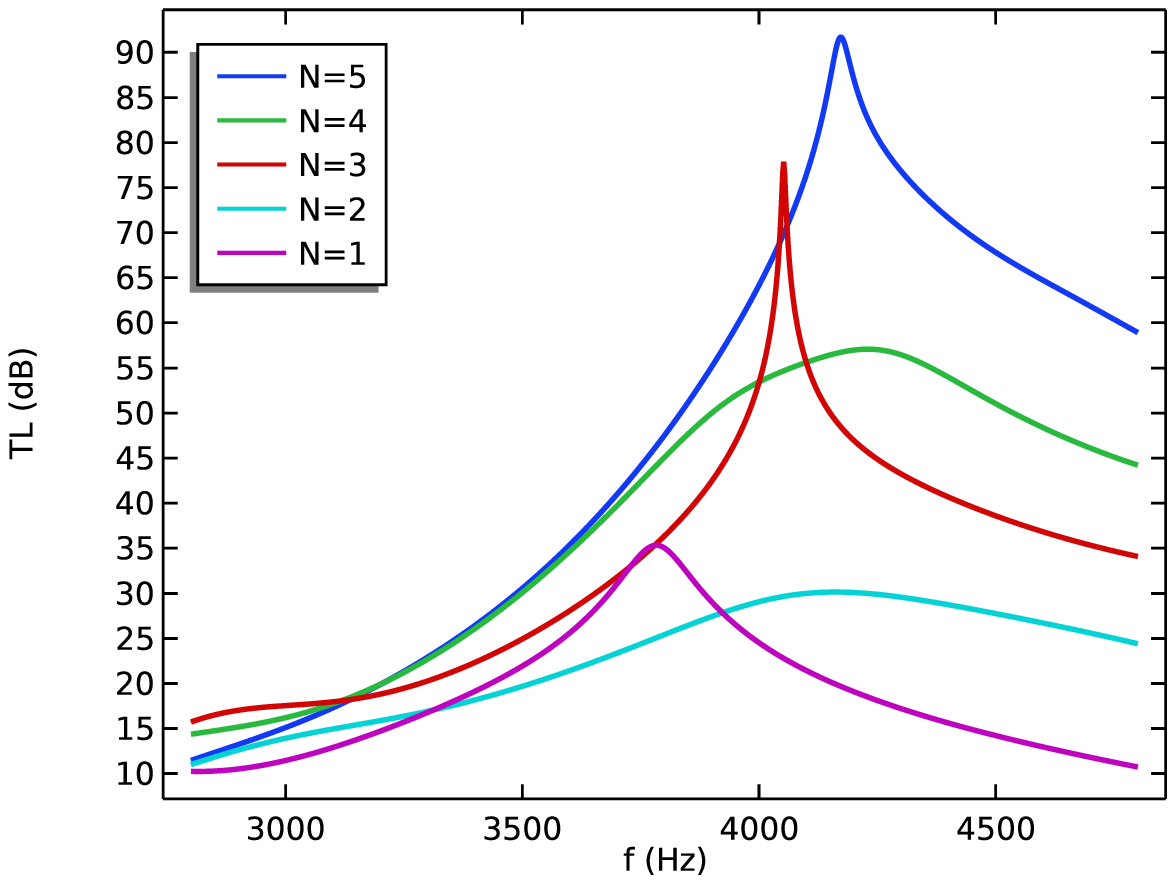}
		\label{TL020}
	}\
	\subfigure[Ma=0.22, N=5]
	{ \includegraphics[scale=0.65]{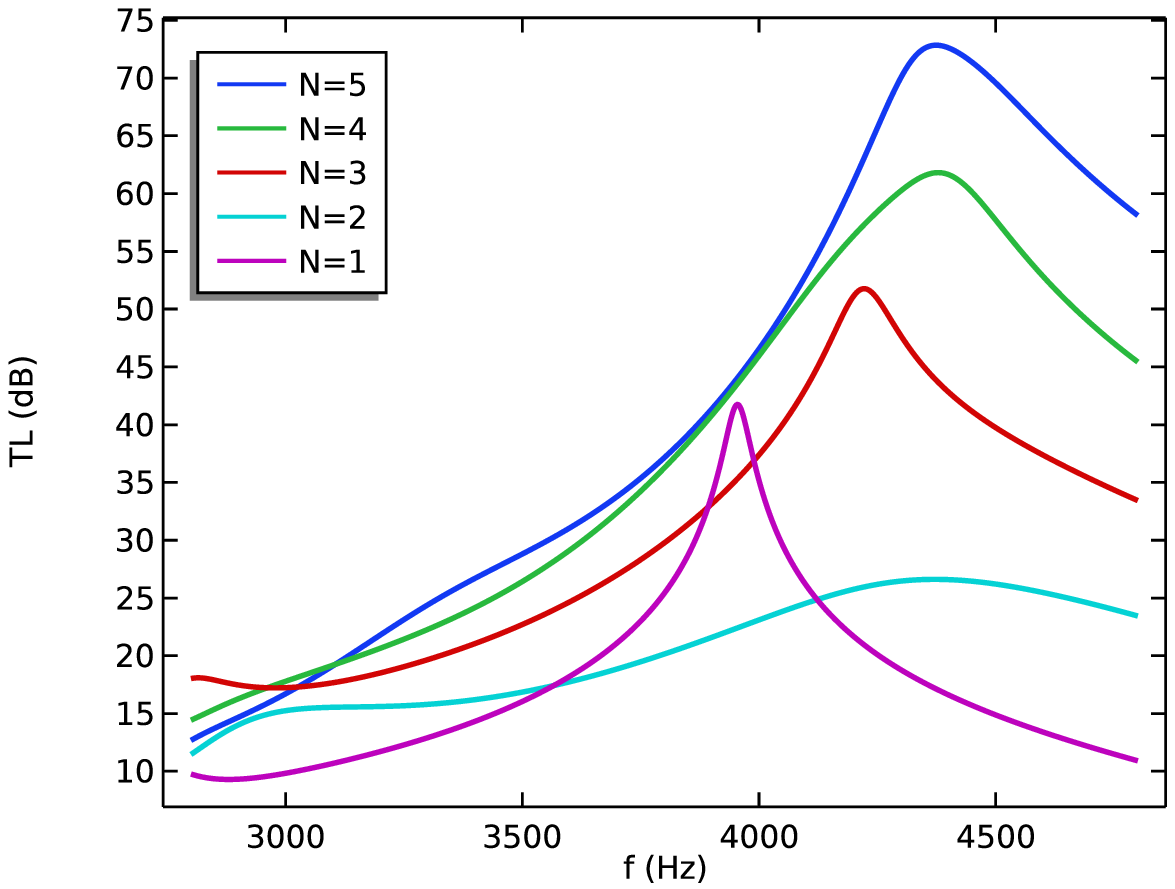}
		\label{TL022}
	}\
	\subfigure[Ma=0.15, N=7]
	{ \includegraphics[scale=0.65]{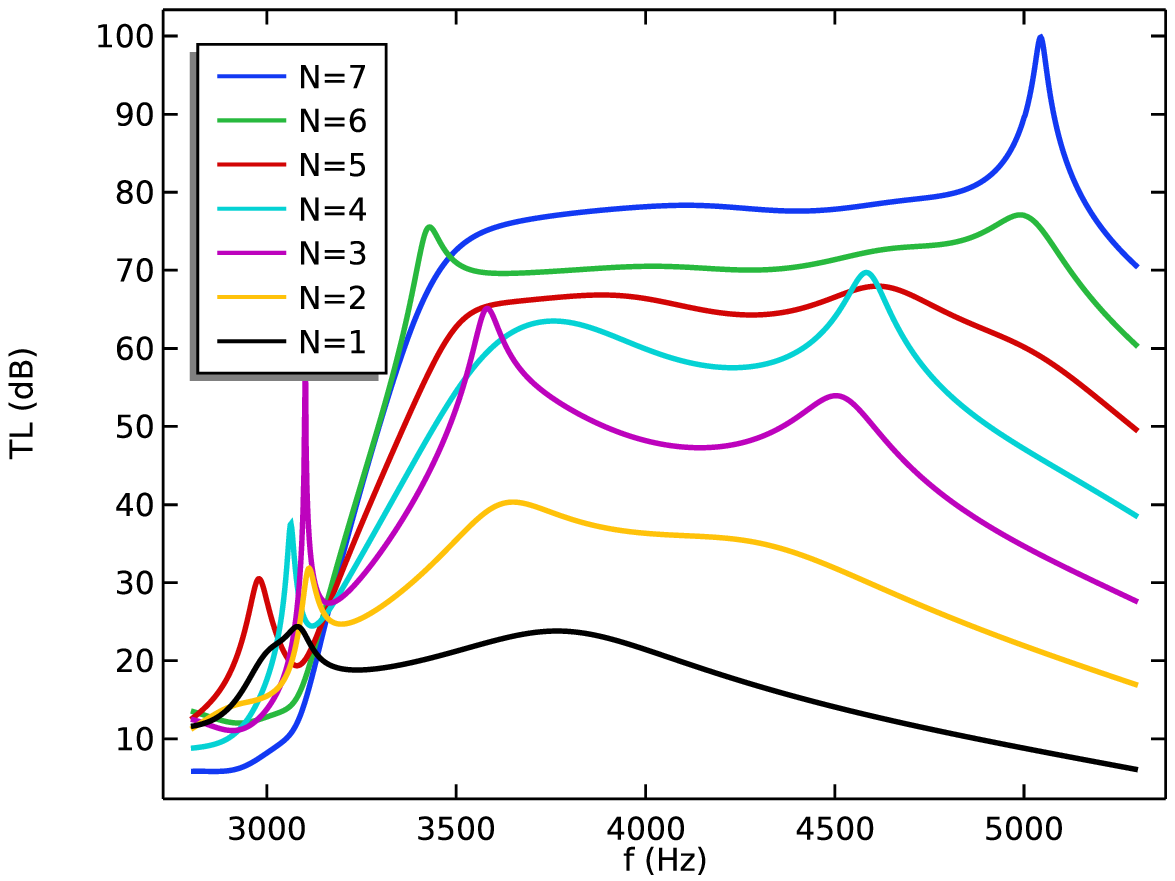}
		\label{TL015_7}
	}\
	
	\caption{ Sound transmission loss corresponding to the size of Ma and the number of annular cavities with $r_a$=25mm in the case of Ma=0.17, 0.20, and 0.22. (* (d) Transmission loss graph investigated while changing the number of annular cavities from 1 to 7 when Ma = 0.15)}
\end{figure}

Fig. 6 and Fig. 7 show the sound transmission loss results calculated while changing the size of Ma and the number of annular cavities in the case of $r_a$=25mm. In order to evaluate the sound transmission loss corresponding to the flow velocity and the number of annular cavities in detail, in Fig. 6(a)-6(d) and Fig. 7(a)-7(c), the number of annular cavities was changed from 1 to 5 and the size of Ma was changed from 0.02 to 0.22. Also, in Fig. 7(d), a detailed calculation of the transmission loss graph is shown while changing the number of annular cavities from 1 to 7 when the flow velocity is Ma = 0.15.

In the case of Ma=0 and $r_a$=25mm, the resonant frequency is 3804.8Hz. If you analyze Fig. 6(a) in detail, you can see that the resonance frequency is rather lower than in the case of no flow. In this case, since Ma=0.02, the convective velocity is small, on the contrary, the Kolomogorov's scale is very large (see Fig. 4). Therefore, the effect of eddy is greater than the effect of convective velocity on sound transmission. The fact that the resonance frequency in the case of N=1,2,3,4 in Fig. 6(a) is lower than that in the case of no flow shows that the effect of eddy shifts the resonance frequency to a lower frequency. The figure also reflects the fact that as N increases, the intrinsic interaction of acoustic metamaterial becomes stronger, and the peak intensity gradually increases. When N=5, a series of changes occurs in the sound transmission loss values. After the first peak of 3677.9Hz occurs, the values gradually change to the second peak of 3938.3Hz, resulting in a widening of the resonance peak. When N=5, the number of annular cavities increases compared to N=1,2,3,4. Therefore, the interaction between the particle vibration in the sound propagation direction in the circular duct and the particle vibration perpendicular to the sound propagation direction in the annular cavity becomes strong. In this case, not only the effect of eddy, but also the intrinsic interaction of acoustic metamaterial has a considerable influence on the sound propagation. Of course, the fact that the largest peak frequency in the figure has been lowered to 3677.9Hz shows that the eddy effect still plays a large role at this time. However, the newly revealed widening properties in the figure reflect that the intrinsic interaction of acoustic metamaterials has a significant effect on sound propagation. In the case of Fig. 6(b), as Ma increases, the effect of convection becomes larger than that in the case of Ma=0.02. This can be known well by seeing that the peak frequency shifted to 3824.7Hz when Ma=0.05 and N=5. However, in this case, the widening property is not significantly different from the case of Ma=0.02. In the case of Fig. 6(c), the effect of convection becomes stronger, and the widening property appears from N=4. In the case of Ma = 0.15 and Ma = 0.17, the intrinsic interaction effect due to the geometry of the acoustic metamaterial, the effect of convection, and the effect of eddy become similar. As a result, in Fig. 6(d) and Fig 7(a), widening property appears from N=1. However, in the case of Ma = 0.17, the effect of convection was greater than in the case of Ma = 0.15, and the peak frequencies were shifted to a larger frequency. When Ma is continuously increased and Ma = 0.20, 0.22 is reached, the effect of convection becomes much larger than the effect of eddy and the geometrical interaction effect of acoustic metamaterial, so the widening property does not appear and the peak frequency moves to a larger frequency (Figure 7(b),(c)). Fig. 7(d) is a graph that investigates the sound transmission loss while increasing the number of annular cavities in the velocity range of Ma = 0.15 where the three interactions are similar. Fig. 7(d) shows that in the case of Ma = 0.15, the geometric interaction effect of acoustic metamaterial begins to play a leading role only when N is 7 or more.

\begin{figure}[hbt!]
	\centering
	\includegraphics[width=.65\textwidth]{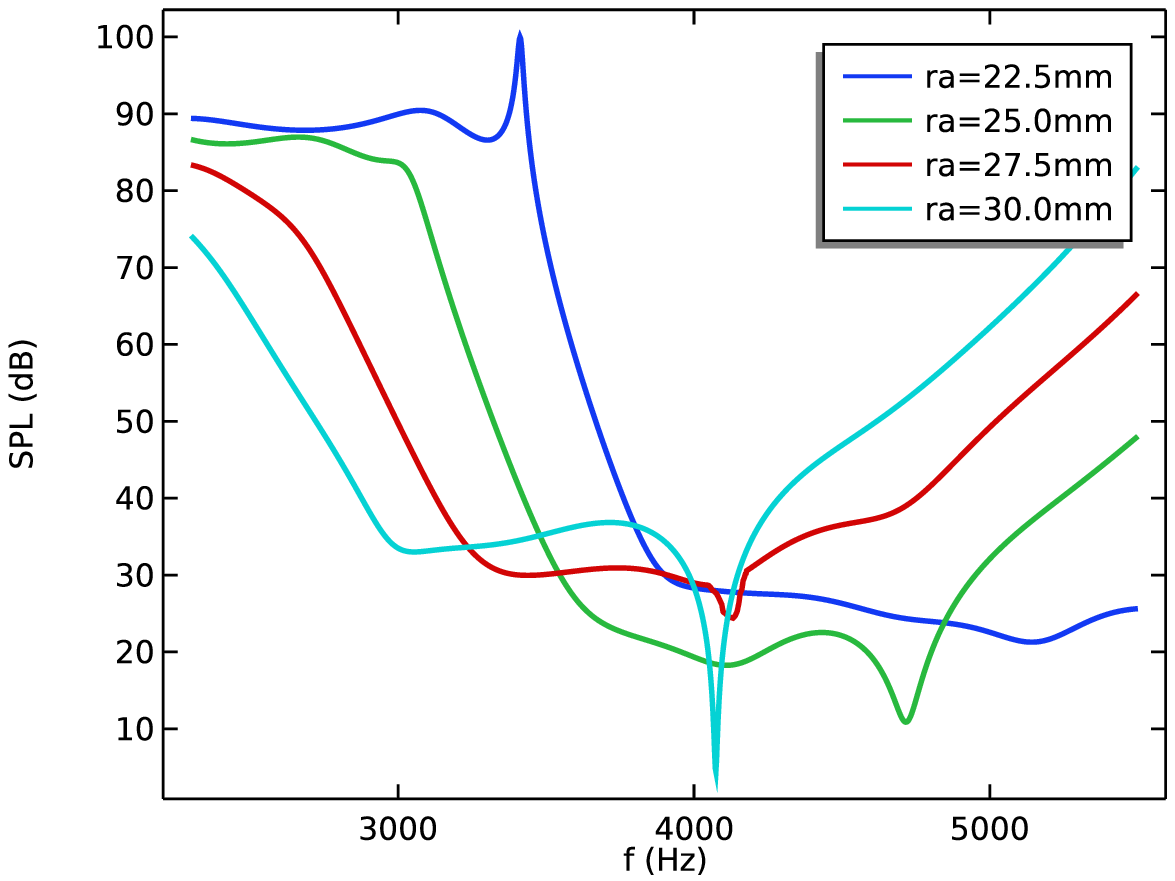}
	\caption{ Sound pressure level results for the radii of annular cavity in case Ma = 0.15.}
\end{figure}

Fig. 8 shows the sound pressure level results for the radii of the annular cavity when Ma = 0.15. In the case of Ma=0.15, it is affected by convection and eddy, but as in the case of Ma=0, as the radius increases, the resonance frequency for the sound pressure level decreases. This fact shows that the particle vibration in the annular cavity analyzed when Ma = 0 is a mechanism that cannot be ignored even in the acoustic propagation analysis with turbulent flow.

\section{Conclusion}
In this study, we discussed the influence of the geometric structure, convection, and eddy on sound propagation in acoustic metamaterial with turbulent flow. First, in order to evaluate the influence of the geometric structure, the acoustic pressure of the acoustic metamaterial was investigated in the case of no flow. Looking at the sound pressure levels corresponding to the frequencies of the sound wave, it has resonance properties, and as the number of annular cavities increases, the resonance strength becomes stronger. Also, as the radius of the annular cavity increases, the resonant frequency decreases. This shows that the sound propagation properties of acoustic metamaterial are closely related to the geometry.

To discuss the problem of sound propagation when turbulent flows into the acoustic metamaterial, not only the effect of convection and eddy, but also the effect of the intrinsic interactions reflecting the geometric structure of the acoustic metamaterial must be considered. Here, the intrinsic interaction refers to the interaction between the particle vibration in the sound propagation direction in the circular duct and the particle vibration perpendicular to the sound propagation direction in the annular cavity. It is not an easy matter to interpret this because all three effects contribute sound transmission. Of course, both convective and eddy effects reduce the intrinsic interaction of acoustic metamaterial in the circular duct, thus reducing the resonant peak intensity for sound transmission. However, considering that the direction of convection is the same as the direction of sound propagation and the direction of eddy is opposite to the direction of sound propagation, the effect of convective flow is opposite to that of eddy. In other words, the convection effect moves the resonance peak in the direction where the frequency is large, and the eddy effect moves the resonance peak in the direction where the frequency is low.  However, when these are combined with the unique interaction properties of the acoustic metamaterial, the widening property of the resonance peak appears. In short, when all three effects cannot be ignored, competition phenomena appear with each other, and as a result, the resonance peak widens and the intensity decreases. In conclusion, the effects of convection, eddy, and intrinsic interactions arising from the unique geometry of acoustic metamaterial appear as the shift of the resonant frequency and the widening properties of the resonant peak.

By using the shift of the resonant frequency and the widening property of the resonant peak studied here, even when turbulence flows, it is possible to block noise by properly controlling the geometric size and shape of the acoustic metamaterial. In particular, this can be used to block noise in transport systems such as train, car, and ship.

\section*{Acknowledgments}
It is a pleasure to thank Un Chol Ri, Yong Kwang Jong and Chol Su Ri for useful discussions. This work was supported by the National Program on Key Science Research of Democratic People's Republic of Korea (Grant No. 20-15-5).

\bibliography{ref_imawc}

\end{document}